# Chromium single photon emitters in diamond fabricated by ion implantation

Igor Aharonovich[a], Stefania Castelletto, Brett C. Johnson, Jeffrey C. McCallum, David A. Simpson, Andrew D. Greentree and Steven Prawer

School of Physics, University of Melbourne, Victoria, 3010, Australia

**Abstract**

Controlled fabrication and identification of bright single photon emitters is at the heart of quantum optics and materials science. Here we demonstrate a controlled engineering of a chromium bright single photon source in bulk diamond by ion implantation. The Cr center has fully polarized emission with a ZPL centered at 749 nm, FWHM of 4 nm, an extremely short lifetime of ~1 ns, and a count rate of $0.5\times10^6$ counts/s. By combining the polarization measurements and the vibronic spectra, a model of the center has been proposed consisting of one interstitial chromium atom with a transition dipole along one of the <100> directions.

(a) Corresponding author i.aharonovich@pgrad.unimelb.edu.au

Quantum technologies require new platforms and paradigms for their efficient fabrication and use. Single photon sources exemplify this need,[1] with their development pushing the bounds of existing materials and fabrication techniques. These sources are the main building blocks for quantum applications, such as quantum communications,[2] quantum metrology[3] and quantum computation. Quantum dots[4] and single molecules[5] have been utilized to generate single photon emission, however a promising class of room temperatures single photon emitters (SPEs) are diamond color centers.

Color centers in monolithic single crystal diamond are strong candidates for scalable quantum information processing (QIP) compared with those in nanocrystals, and hence several designs for large scale quantum architecture employing monolithic diamond have been proposed.[6,7]

The nitrogen-vacancy (NV) center is the most studied center for QIP applications owing to the availability of optical read-out of the individual electronic spin state,[8,9] leading to advanced applications such as nanomagnetometry.[10,11] However, the NV center has fundamental limitations - strong phonon coupling of the excited state which results in a broad photoluminescence (PL) spectrum (~ 100 nm) of which the zero phonon line (ZPL) makes up ~ 4%. Emission of single photons in the ZPL is then extremely weak, typically on the order of a few thousands of photons per second. Such count rates are insufficient for the realization of advanced QIP protocols. These constraints motivate the need to identify and fabricate improved diamond based SPEs.

To move beyond the limitations of the NV center, alternative diamond based SPEs[12,13] such as the silicon vacancy (Si-V),[14] or the nickel related complex (NE8)[15] have been investigated. However, the very low emission rates of the Si-V center, and the difficulty to controllably fabricate both of these centers at the single center level, has motivated the search for alternative SPEs. A major challenge in the material science community in this search is to unambiguously identify new diamond based color centers and determine reproducible techniques to fabricate such sources in both bulk single crystal diamond and nanodiamonds (NDs).

Circumstantial evidence of Cr-related centers in nanodiamond grown on sapphire substrates has been recently reported,[16] however an unambiguous determination of its composition and structure is yet to be ascertained. To explore the origins of novel color centers and their fundamental physical properties, the fabrication of these centers in monolithic single crystal diamond, rather than nanocrystals is necessary. Ion implantation[17] offers the unique advantage generating optical centers based solely on the constituents used in the implantation process along with the known and well characterized elements present in the bulk diamond. Implantation is therefore particularly advantageous when trying to identify the origin of an unknown color center.

Here we present and verify a new class of single photon emitters engineered in monolithic single crystal diamond by the co-implantation of chromium and oxygen atoms. The ultra bright emission in the near infrared (NIR) at 749 nm and short excited state lifetime (~1 ns) surpass the performance of previously identified SPEs in single crystal (bulk) diamond. Cr-related luminescence has been identified in sapphire[18] and silicon[19]. While the origin of the luminescence in silicon is still under investigation,[20] the atomic structure of the center in sapphire is attributed to a substitutional $Cr^{3+}$ ion. With this in mind, the co-implantation of Cr and O atoms into single crystal diamond was performed to modify the oxidation state of the chromium in the diamond lattice.

CVD single crystal diamond samples (type IIA, [N] < 1 ppm, [B] < 0.05 ppm, Element Six) were co-implanted with 50 keV chromium and 19.5 keV oxygen to a fluencies of $1\times10^{11}$ $Cr/cm^2$ and $1.5\times10^{11}$ $O/cm^2$, respectively. The implantation energies were chosen to maximize the proximity of the two atomic species. Fig. 1a shows a Monte Carlo simulation using the Stopping Range of Ions in Matter [SRIM] software of implanted Cr and O atoms accelerated to 50 keV and 19.5 keV, respectively. The projected range of these implantations was approximately 25 nm below the diamond surface which allowed efficient out-coupling of the light emitted from the center. After the implantation, the diamonds were annealed at 1000°C in a forming gas ambient (95%-Ar-5%$H_2$) for one hour. Note that the annealing step applied after the implantation (1000°C) is not sufficient to cause any diffusion of Cr atoms in the stiff diamond lattice;[21,22] hence the formed centers are indeed located in the depth of the implantation.

Fig. 1b shows a fluorescence confocal map recorded from the (100) oriented single crystal diamond after the co-implantation of chromium and oxygen. The bright spots in the confocal image correspond to the engineered centers from the ion implantation process.

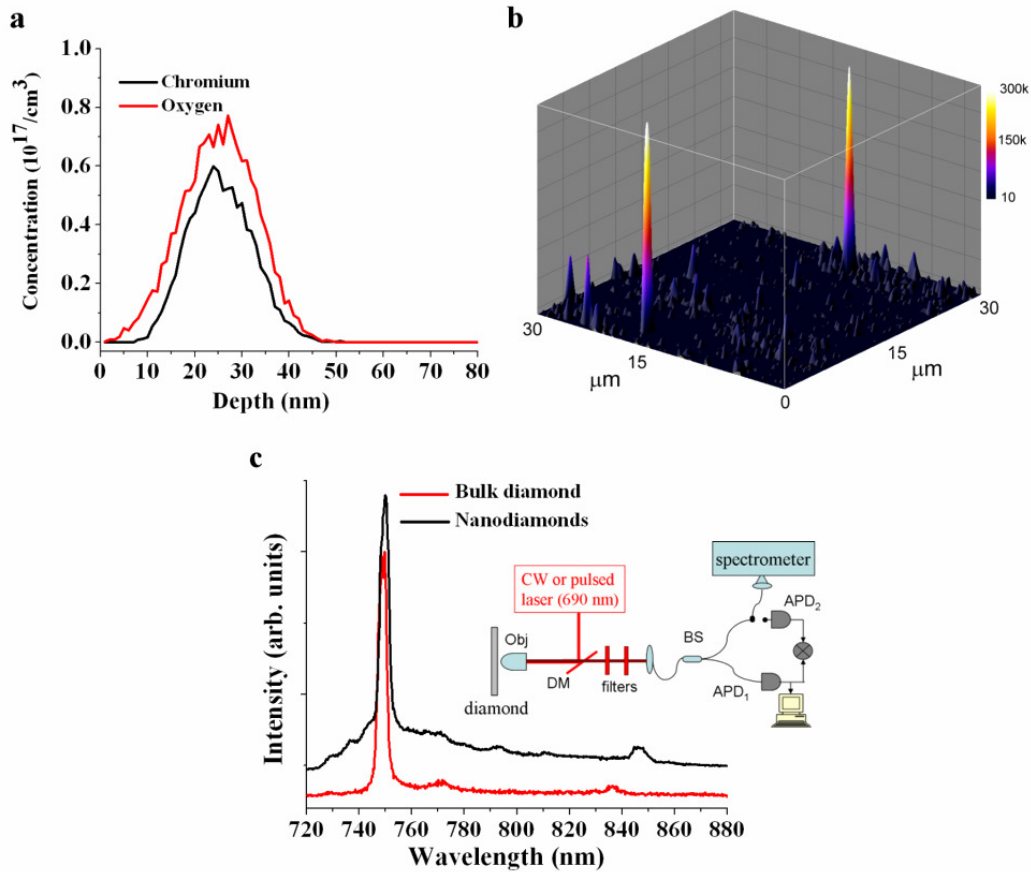

**Fig 1. (Color online) Chromium implanted diamond. (a)** Monte Carlo simulation of Cr and O distribution inside a diamond following 50 keV Cr and 19.5 keV O implantation. **(b)** A confocal map recorded from the diamond sample co-implanted with $1\times10^{11}$ Cr/cm$^2$ and $1.5\times10^{11}$ O/cm$^2$. The bright spots correspond to Cr centers in diamond **(c)** PL spectrum recorded at room temperature from a bright spot on the confocal map shown in Fig 1 (red curve). The black curve is the PL spectrum recorded from nanodiamonds grown on sapphire following the procedure reported elsewhere.[15] The data from the nanodiamonds was displaced vertically for clarity. Inset, is the optical setup.

Fig. 1c shows a narrow PL spectrum, recorded at room temperature using a continuous wave (CW) diode laser emitting at 682 nm from one of the bright spots on the confocal map shown in Fig. 1a (red curve), with a ZPL centered at 749 nm and a full width at half maximum (FWHM) of 4 nm. Similar PL lines were consistently observed from different bright spots across the sample. The co-implantation technique yielded 2-4 Cr centers with a ZPL centered at 749 nm per 50 μm², resulting in a conversion efficiency of less than 0.1%. The black curve in Fig. 1c shows as a comparison a PL spectrum recorded from a center in nanodiamond grown on a sapphire substrate.[16] The excellent agreement between the PL spectra, in terms of the ZPL and FWHM, in bulk diamond and NDs clearly indicates that these class of emitters can be attributed to Cr.

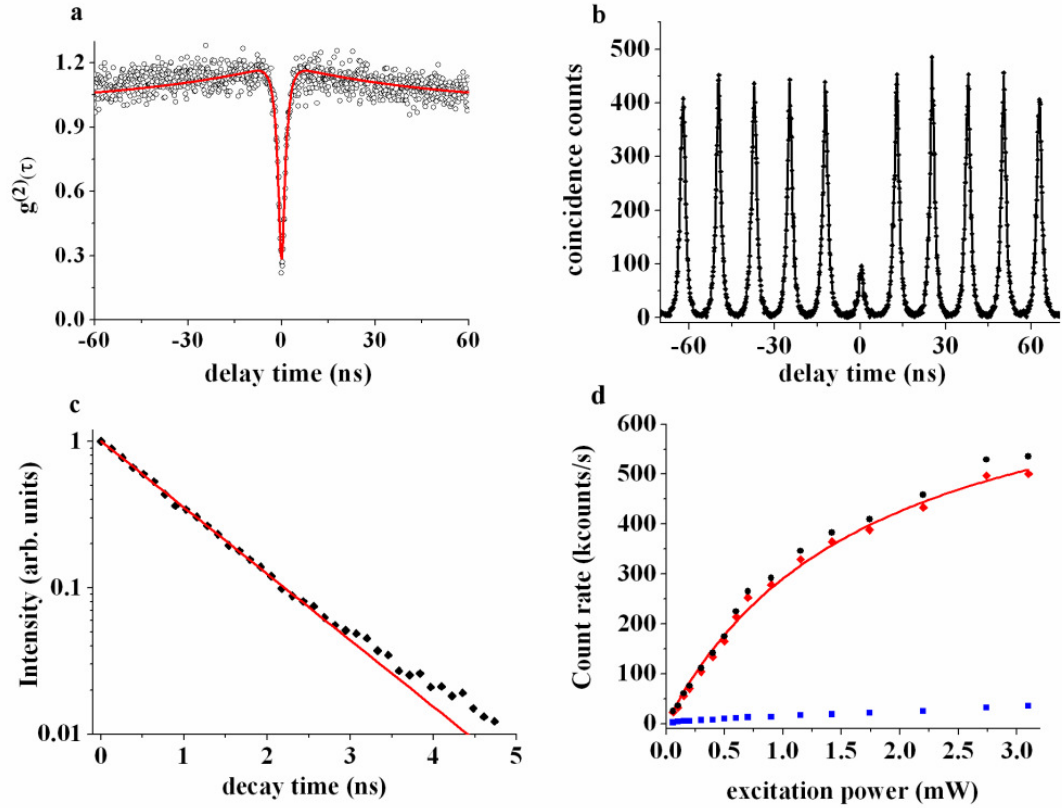

**Fig. 2. (Color online) Optical characterization of the Cr center. (a) A normalized second-order correlation function, $g^{(2)}(\tau)$, corresponding to the PL line shown in Fig. 1c, demonstrating single photon emission ($g^{(2)}(0)=0.2$). The bunching ($g^{(2)}(\tau)>1$) indicates of a presence of a third metastable state. The red line is the theoretical fit based on a three level model, taking into account the time response of the detectors. (b) Triggered SP emission is verified by exciting with a pulsed laser at 80 MHz. The deviation from zero of the $g^{(2)}(0)$ under CW excitation is attributed to the background and the time response jitter of the photo detectors and correlation electronics, while in the pulsed regime the deviation from zero can be attributed to background due to non perfect polarization condition of the pulsed excitation laser. (c) A direct lifetime measurement (dots) with a fit of a single exponential**

(red curve). A lifetime of 1±0.1 ns is deduced from the fit. (d) Single photon emission count rate recorded from the same emitter as a function of excitation power. The blue squares represent the background noise; the black circles represent the raw data and the red rhombs represent the background corrected count rate. The fit (red curve) was obtained from the solution of a three level system.

The photon statistics of the PL in single crystal diamond were studied by recording the second-order correlation function, $g^{(2)}(\tau) = <I(t)I(t+\tau)>/<I(t)>^2$, using a fibre-based Hanbury Brown & Twiss setup[23]. Fig. 2a shows the corresponding, $g^{(2)}(\tau)$ function, for the 749 nm center. The dip at zero delay time $g^{(2)}(\tau=0)=0.2$ is characteristic of non classical light and verifies single photon emission. The deviation from zero at $\tau=0$ is attributed to the background within the single crystal diamond. The photon bunching ($g^{(2)}(\tau)>1$) observed at longer delay times is indicative of a three level system with a long lived metastable state.[8, 15] To verify single photon emission on demand, a key feature for many quantum optical applications, the center was excited with a 690 nm pulsed laser with a repetition rate of 80 MHz, as shown in Fig. 2b. The vanishing peak at $\tau=0$ indicates that only one photon is emitted per excitation pulse. The excited state lifetime of the same emitter is shown in Fig. 2c and was obtained under pulsed excitation at 40 MHz. From the single-exponential fit to the fluorescence decay, an excited state lifetime of 1.0±0.1 ns is deduced. This is a short fluorescence lifetime when compared to other known SP emitters in diamond,[8,15] indicative of a very strong dipole moment in the radiative transition, which we have determined to be 60 Debye. The measured lifetime is in agreement with the one measured for the center in nanodiamonds (1.1±0.1 ns).

To quantify the single photon efficiency, the count rate was measured as a function of the excitation power. Fig. 2d shows the measured single photon count rate as a function of excitation power. A measured count rate of ~0.5×10$^6$ counts/s makes the Cr center the brightest reported single photon source in a monolithic single crystal diamond to date and within an order of magnitude of the brightest nanodiamond sources.[12,16] Furthermore, when these centers are integrated with the recent diamond based antennas,[24] even higher count rates can be expected. Note that the center is located only a few tens of nm below the diamond surface which is advantageous for efficient coupling of the emitted light to external cavities or waveguides.[25]

To understand the dipole orientation of the Cr center relative to the crystal axis, excitation polarization measurements on the emitters with a ZPL centered at 749 nm were performed[26,27]. Prior to each measurement, the antibunching was recorded to verify that only single emitters were addressed. All single 749 nm centers studied here were found to exhibit a dependence on the incident excitation polarization. Two distinct polarization dependencies were identified by characterizing numerous single 749 nm emitters and these are shown in Fig. 3a for four separate single emitters. By varying the incident polarization angle by 90 degrees in the respective configuration the PL intensity evolves from maximum to minimum intensity, with extinction ratios of 96%, as seen in Fig. 3a. This is consistent with the characteristic dipole transition behavior.

Given that the diamond sample is (100) oriented, the direction of the center transition dipole moment may be determined. The electric field of the incoming laser is parallel to the (100) plane of the crystal. Since the modulation of the excitation polarization angle results in an absolute extinction of the PL, the dipole of the center must be aligned in the [0yz] direction, otherwise full extinction of the PL would not be observed. Furthermore, the period of the minima (or the maxima) of the emitters is 90 degrees, which indicates that only two different projections of the dipole on the (100) plane are possible. Due to the diamond structure, the only crystallographic directions that meet these conditions are the [010] or [001]. Note that the dipole of the center can be also aligned along the [100] direction, however, in this case it will always be perpendicular to the excitation electric field and thus, will never be excited. Thus, the transition dipole moment of the observed centers is aligned in one of the <100> directions.

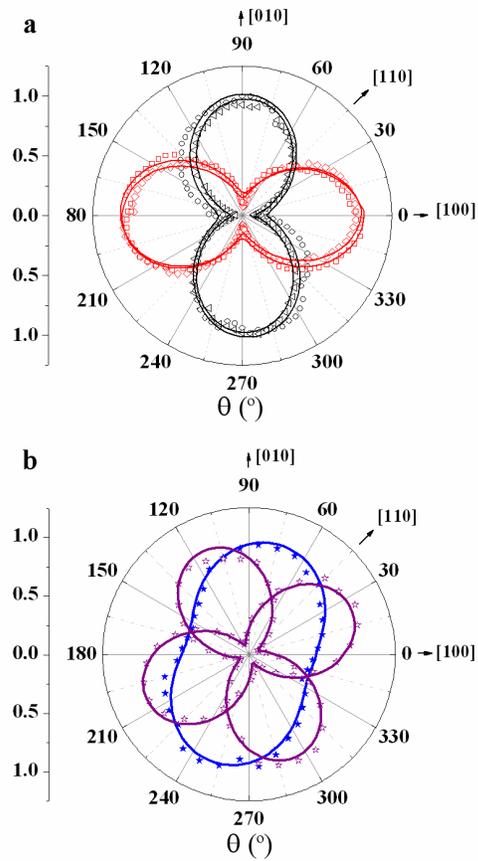

**Fig. 3. (color online) Polarization properties and the structure of the Cr center. (a) Polarization measurements of the 749 nm emitter represented in polar coordinates. θ is the polarization angle referred to the main crystallographic axes. PL intensity versus polarization of the excitation laser for four different SP emitters (black triangles, black circles, red squares and red diamonds) with a ZPL centered at 749 nm. An extinction of the signal down to the typical background level is demonstrated. A period of nearly 90 degrees between the maximum and the minimum of different emitters (black and red colors) is observed. (b) Modulation of the emission intensity measured by rotating a linear polarization analyzer at the detection channel, with a fixed polarization of the excitation field (blue stars). By introducing a quarter wave plate before the analyzer, full polarization contrast was achieved. Purple stars show two possible polarization emissions for two different wave plate positions. (c) A schematic model of the Cr center. The Cr atom (pink) is occupying an interstitial site in the diamond lattice. The dipole of the center is oriented along one of the <100> crystallographic directions.**

The emission polarization of the system was studied by fixing the polarization excitation to maximum and rotating the polarizer at the detection channel. The linear polarization dependence (Fig. 3b, blue stars) may be attributed to the fact that the emission and the absorption dipoles are not parallel. To compensate this effect, a quarter wave plate was inserted in the detection channel before the analyzer which resulted in a full extinction of the emitted light (Fig. 3b purple open stars). Different orientations of the modulated emission can be obtained by varying the quarter wave plate position, as shown in Fig. 3b for two different positions. The fully polarized light indicates that transitions such as thermal relaxation within the excited state manifold, do not exist under our experimental conditions and the decay to the ground state occurs from an individual excited state level. This is an important characteristic for the generation of indistinguishable single photons and quantum entanglement[28].

Finally, we discuss the atomic structure of the Cr center in diamond. A substitutional atom in a diamond lattice exhibits tetrahedral symmetry and would possess a different excitation polarization dependence than the observed here. Furthermore, most of the substitutional atoms in the diamond structure form complexes aligned along the <111> axis (e.g. substitutional nitrogen adjacent to a vacancy). As the excitation polarization measurements in this work result in a complete extinction of the PL and indicate that the excitation dipole of the center is located on the {001} planes, the incorporation of Cr atoms as substitutional is excluded. Similarly, an adjacent vacancy, which generally occupies the (111) location site is also excluded. Since the size of the Cr atom is significantly larger than the carbon matrix, an interstitial location is most stable. We therefore conclude that the Cr atom occupies an interstitial related site, with a possible distortion of a surrounding diamond lattice.

To quantify the number of Cr atoms forming the color centre we investigate the effects of local vibrations due to the number of atoms forming the defect. A defect consisting of an impurity atom of mass $M_I$ located in a matrix with host atom mass $M_C$, generates quasilocal vibrations which can be described by the following equation: $\omega_{QL} = \omega_D \sqrt{M_C/3(kM_I - M_C)}$ where $\omega_D = 150\,\text{meV}$ is the Debye frequency of the diamond lattice and $k$ is the number of impurity atoms involved in the quasilocal vibration[29]. For one Cr atom, the vibration should occur 47 meV from the ZPL, which in the present case occurs at 771 nm. Following the measured PL from the bulk diamond presented in Fig. 1c, there is good agreement with the theory as a vibronic feature can be observed at 771 nm, which corresponds to 47 meV shift from the ZPL.

This reasoning, along with the magnitude of the dipole moment discussed above, allows us to propose a model of the Cr center. One interstitial Cr atom in the diamond lattice with a dipole moment oriented parallel or perpendicular to one of the <001> crystallographic axis. Following this model the center belongs to the $D_2d$ group symmetry. We propose the name UM2 for this center.

We can not at this stage unambiguously identify the role of the oxygen in the co-implantation except to note that the co implantation significantly enhances the formation of the UM2 centre; one may assume that its behavior as an electron trap may set the charge state of the chromium. A test implantation of chromium only into the same type of single crystal diamond, followed by the same annealing treatment yielded only limited PL lines centered at 749 nm. The ratio of UM2 centers observed in the co-implantation of chromium and oxygen compared to the chromium only implantation was ~ 10:1.

In summary, we demonstrate, for the first time, the controlled engineering of a chromium bright single photon source in bulk diamond, labeled as UM2. The center located in close proximity to the diamond surface (~25 nm), has fully polarized emission with a ZPL centered at 749 nm, FWHM of 4 nm, an extremely short lifetime of 1 ns, and exhibits a count rate of approximately $0.5 \times 10^6$ counts/s, making it the brightest SPE in bulk diamond to date. By combining the polarization measurements and the vibronic spectra, a model of the center has been proposed consisting of one interstitial chromium atom belonging to the $D_2d$ group symmetry and exhibits a transition dipole along one of the <100> directions. The enhanced properties of this centre compared with the known diamond centers and the scalable fabrication technique, suggest that the UM2 center will be important for a range of applications including quantum metrology, quantum cryptography and biomarking.

**Acknowledgements**

The authors thank Faruque Hossain and Fedor Jelezko for many helpful discussions. This work was supported by the Australian Research Council, The International Science Linkages Program of the Australian Department of Innovation, Industry, Science and Research (Project No. CG110039) and by the European Union Sixth Framework Program under the EQUIND IST-034368. ADG is the recipient of an Australian Research Council Queen Elizabeth II Fellowship (project No. DP0880466).